\documentclass[12pt] {article}
\usepackage{psfrag}
\usepackage{graphicx}
\usepackage{latexsym,amsfonts}
\usepackage{amssymb}
\begin{document}
\setcounter{page}{1}
\def\theequation{\arabic{section}.\arabic{equation}}
\def\theequation{\thesection.\arabic{equation}}
\setcounter{section}{0}

\title{Quantum field theory of a free massless (pseudo)scalar field in
1+1--dimensional space--time as a test for the massless Thirring model
}

\author{M. Faber\thanks{E--mail: faber@kph.tuwien.ac.at, Tel.:
+43--1--58801--14261, Fax: +43--1--5864203} ~~and~
A. N. Ivanov\thanks{E--mail: ivanov@kph.tuwien.ac.at, Tel.:
+43--1--58801--14261, Fax: +43--1--5864203}~\thanks{Permanent Address:
State Polytechnic University, Department of Nuclear Physics, 195251
St. Petersburg, Russian Federation}}

\date{\today}

\maketitle
\vspace{-0.5in}
\begin{center}
{\it Atominstitut der \"Osterreichischen Universit\"aten,
Arbeitsbereich Kernphysik und Nukleare Astrophysik, Technische
Universit\"at Wien, \\ Wiedner Hauptstr. 8-10, A-1040 Wien,
\"Osterreich }
\end{center}

\begin{center}
\begin{abstract}
We analyse different approaches to the description of the quantum
field theory of a free massless (pseudo)scalar field defined in
1+1--dimensional space--time which describes the bosonized version of
the massless Thirring model. These are (i) axiomatic quantum field
theory, (ii) current algebra and (iii) path--integral. We show that
the quantum field theory of a free massless (pseudo)scalar field
defined on the class of the Schwartz test functions ${\cal
S}_0(\mathbb{R}^{\,2})$ connects all these approaches. This quantum
field theory is well--defined within the framework of Wightman's
axioms and Wightman's positive definiteness condition. The physical
meaning of the definition of Wightman's observables on the class of
test functions from ${\cal S}_0(\mathbb{R}^{\,2})$ instead of ${\cal
S}(\mathbb{R}^{\,2})$, as required by Wightman's axioms, is the
irrelevance of the collective zero--mode related to the collective
motion of the ``center of mass'' of the free massless (pseudo)scalar
field, which can be deleted from the intermediate states of
correlation functions (Eur. Phys. J. C {\bf 24}; 653 (2002)).  In such
a theory the continuous symmetry, induced by shifts of the massless
(pseudo)scalar field, is spontaneously broken and there is a
non--vanishing spontaneous magnetization. The obtained results are
discussed in connection with Coleman's theorem asserting the absence
of Goldstone bosons and spontaneously broken continuous symmetry in
quantum field theories defined in 1+1--dimensional space--time.
\end{abstract}
\end{center}

\newpage

\section{Introduction}
\setcounter{equation}{0}

\hspace{0.2in} In 1964 Wightman has delivered his seminal lectures
[1], where he has formulated his point of view concerning the quantum
field theory of a free massless (pseudo)scalar field $\vartheta(x)$,
which can be described by the Lagrangian
\begin{eqnarray}\label{label1.1}
{\cal L}(x) =
\frac{1}{2}\,\partial_{\mu}\vartheta(x)\partial^{\mu}\vartheta(x),
\end{eqnarray}
invariant under the continuous Abelian symmetry group induced by field
shifts
\begin{eqnarray}\label{label1.2}
\vartheta(x) \to \vartheta\,'(x) = \vartheta(x) + \alpha
\end{eqnarray}
with the parameter $\alpha \in \mathbb{R}^{\,1}$. 

For the definition of the quantum field theory, described by the
Lagrangian (\ref{label1.1}), Wightman introduced observables defined
by
\begin{eqnarray}\label{label1.3}
\vartheta(h) = \int d^2x\,h(x)\,\vartheta(x),
\end{eqnarray}
where $h(x)$ are test functions from the Schwartz class ${\cal
S}(\mathbb{R}^{\,2})$ or ${\cal S}_0(\mathbb{R}^{\,2}) = \{h(x) \in
{\cal S}(\mathbb{R}^{\,2}); \tilde{h}(0) = 0\}$ [1]. The function
$\tilde{h}(k)$ is the Fourier transform of $h(x)$ defined by
\begin{eqnarray}\label{label1.4}
\tilde{h}(k) = \int d^2x\,h(x)\,e^{\textstyle + ik\cdot x}\quad,\quad
h(x) &=& \int \frac{d^2k}{(2\pi)^2}\,\tilde{h}(k)
\,e^{\textstyle - ik\cdot x}.
\end{eqnarray}
In terms of Wightman's observable $\vartheta(h)$ one can define a
quantum state $|h\rangle$ 
\begin{eqnarray}\label{label1.5}
|h\rangle = \vartheta(h)|\Psi_0\rangle
\end{eqnarray}
with the norm $||h||$ given by
\begin{eqnarray}\label{label1.6}
||h||^2 &=& \langle h|h\rangle = \langle
  \Psi_0|\vartheta(h)\vartheta(h)| \Psi_0\rangle = \int\!\!\!\int
  d^2xd^2y\,h^*(x)\,\langle
  \Psi_0|\vartheta(x)\vartheta(y)|\Psi_0\rangle\,h(y) = \nonumber\\
  &=&\int\!\!\!\int d^2xd^2y\,h^*(x)\,D^{(+)}(x-y; \mu)\,h(y).
\end{eqnarray}
Here $|\Psi_0\rangle$ is a vacuum state and $D^{(+)}(x-y; \mu)$ is the
Wightman function defined by [1]
\begin{eqnarray}\label{label1.7}
\hspace{-0.3in}&&D^{(+)}(x -y ;\mu) = \langle
\Psi_0|\vartheta(x)\vartheta(y)|\Psi_0\rangle =\nonumber\\
\hspace{-0.3in}&&=
\frac{1}{2\pi}\int^{\infty}_{-\infty}\frac{dk^1}{2k^0}\,e^{\textstyle
-\,i\,k\cdot (x - y)} = - \frac{1}{4\pi}{\ell n}[-\mu^2(x - y^2 +
i\,0\cdot\varepsilon(x^0 - y^0)],
\end{eqnarray}
where $\varepsilon(x^0 - y^0)$ is the sign function, $(x - y)^2 = (x^0
- y^0)^2 - (x^1 - y^1)^2$, $k\cdot (x - y) = k^0(x^0 - y^0) - k^1(x^1
- y^1)$, $k^0 = |k^1|$ is the energy of a free massless (pseudo)scalar
quantum with momentum $k^1$ and $\mu$ is the infrared cut--off
reflecting the infrared divergence of the Wightman function
(\ref{label1.7}).

Wightman's analysis of a quantum field theory, described by the
Lagrangian (\ref{label1.1}), can be summarized as ``$\ldots$ there is
no such mathematical object as a free field with mass zero in
two--dimensional space--time unless one of the usual assumptions is
abandoned.''  [1]. The usual assumptions are Wightman's axioms and
Wightman's positive definiteness condition [1--4].

The main problem promoting Wightman to make such a strong assertion
was the infrared divergence of the Wightman function (\ref{label1.7}).
Due to this infrared divergence the Wightman function $D^{(+)}(x -
y;\mu)$ does not satisfy Wightman's positive definiteness condition
[1--3]
\begin{eqnarray}\label{label1.8}
||h||^2 = \int\!\!\!\int d^2xd^2y h^*(x) D^{(+)}(x-y;\mu) h(y) \ge 0
\end{eqnarray}
on the Schwartz test functions $h(x)$ from ${\cal
S}(\mathbb{R}^{\,2})$ [1--3]. This can be easily seen in the momentum
representation. Substituting (\ref{label1.4}) and (\ref{label1.7}) in
(\ref{label1.8}) we get
\begin{eqnarray}\label{label1.9}
||h||^2 = \frac{1}{2\pi}\int^{\infty}_{-\infty}
\frac{dk^1}{2k^0}\,|\tilde{h}(k^0, k^1)|^2.
\end{eqnarray}
In the light--cone variables $k_+ = k^0 + k^1$, $k_- = k^0 - k^1$ and
$d^2k = \frac{1}{2}dk_+dk_-$ the r.h.s. of (\ref{label1.9}) reads [5]
\begin{eqnarray}\label{label1.10}
||h||^2 =
\frac{1}{2\pi}\int^{\infty}_0\frac{dk_+}{k_+}\,|\tilde{h}(k_+, 0)|^2,
\end{eqnarray}
where we have used the Fourier transform of the Wightman function
$D^{(+)}(x-y; \mu)$ equal to [5]
\begin{eqnarray}\label{label1.11}
F^{(+)}(k) &=& \int d^2(x-y)\,e^{\textstyle
 +ik\cdot(x-y)}\,D^{(+)}(x-y; \mu) = 2\pi\,\theta(k^0)\,\delta(k^2)
 =\nonumber\\ &=& 2\pi\,\frac{\theta(k_+)}{k_+}\,\delta(k_-) +
 2\pi\,\frac{\theta(k_-)}{k_-}\,\delta(k_+)
\end{eqnarray}
and symmetric test functions $\tilde{h}(k_+, k_-) = \tilde{h}(k_-,
k_+)$ for simplicity.  

If the test functions $h(x)$ belong to the Schwartz class ${\cal
S}(\mathbb{R}^{\,2})$ with $\tilde{h}(0) \neq 0$, the integral over
$k^1$ in (\ref{label1.9}) has a logarithmic divergence. Since this
integral is related to a norm of a quantum state, it cannot be
infinite. Therefore, the integral over $k^1$ should be
regularized. The former can be carried out within the theory of
generalized functions [26]. The regularized expression is equal to
\begin{eqnarray}\label{label1.12}
\hspace{-0.5in}||h||^2 &=& \int\!\!\!\int
d^2xd^2y\,h^*(x)\,D^{(+)}(x-y; \mu)\,h(y)_{R} = \nonumber\\
\hspace{-0.5in}&=&\frac{1}{2\pi}
\int^{\infty}_{-\infty}\frac{dk^1}{2k^0}\,(|\tilde{h}(k)|^2 -
|\tilde{h}(0)|^2\,) = \frac{1}{2\pi}\int^{\infty}_0\frac{dk_+}{k_+}
(|\tilde{h}(k_+, 0)|^2 - |\tilde{h}(0, 0)|^2\,).
\end{eqnarray}
The test functions of the Schwartz class are fast decreasing for $k^1
\to \pm\,\infty$ and correspondingly for $k_{\pm} \to \infty$. This
implies that the momentum integrals of the regularized expression
(\ref{label1.12}) are negative definite. This yields
\begin{eqnarray}\label{label1.13}
||h||^2 =\int\!\!\!\int d^2xd^2y\,h^*(x)\,D^{(+)}(x-y; \mu)\,h(y)_{R}
< 0.
\end{eqnarray}
Hence, Wightman's positive definiteness condition (\ref{label1.8}) is
violated for quantum states in the quantum field theory of a free
massless (pseudo)scalar field described in terms of Wightman's
observables defined on test functions $h(x)$ from the Schwartz class
${\cal S}(\mathbb{R}^{\,2})$ with $\tilde{h}(0) \neq 0$.

According to Wightman ``one way to make the positive definiteness of''
(\ref{label1.8}) ``irrelevant is to restrict the class of test
functions'' from ${\cal S}(\mathbb{R}^{\,2})$ to ${\cal
S}_0(\mathbb{R}^{\,2}) = \{h(x) \in {\cal S}(\mathbb{R}^{\,2});
\tilde{h}(0) = 0\}$ [1].

In this connection the question could be asked: ''Why can quantities
like test functions, which do not enter to the Lagrangian of a free
massless (pseudo)scalar field and, correspondingly, do not affect the
dynamics of this system, play such a crucial role for the existence of
a quantum field theory of a free massless (pseudo)scalar field.

In our recent paper [5] we have made an attempt to understand
Wightman's observables $\vartheta(h)$, defined by (\ref{label1.3}). We
consider the Fourier transform $\tilde{h}(k)$ of the test function
$h(x)$ as an {\it apparatus} function related to the {\it resolving
power} of the device, which the observer uses for the detection of
quanta of a free massless (pseudo)scalar field. Of course, such an
interpretation is also rather questionable. This is due to the
neutrality and decoupling of the quanta of the free massless
(pseudo)scalar field $\vartheta(x)$ from everything. Nevertheless, if
in spite of this we assume a measurability of these quanta, Wightman's
statement, concerning the possibility to define a quantum field theory
of a free massless (pseudo)scalar field $\vartheta(x)$ on the class of
test functions from ${\cal S}_0(\mathbb{R}^{\,2}) = \{h(x) \in {\cal
S}(\mathbb{R}^{\,2}); \tilde{h}(0) = 0\}$, can be interpreted as
impossibility to detect quanta with zero energy and momentum $k^0 =
k^1 = 0$. We call them below {\it zero--mode quanta}.

In our article [6] we have found that zero--mode quanta are related to
the collective shift of the field $\vartheta(x)$ describing the motion
of the ``center of mass''. We have shown that the removal of the
collective zero--mode allows to formulate the quantum field theory of
the free massless (pseudo)scalar field without infrared divergences
[6]. For a free quantum system the exclusion of the collective
zero--mode does not affect the evolution of the system caused by a
relative motion in it [6]. The possibility to remove the collective
zero--mode has been realized within the path--integral approach to the
description of a quantum field theory of a free massless
(pseudo)scalar field $\vartheta(x)$ in terms of the generating
functional of Green functions $Z[J]$ of the free massless
(pseudo)scalar field $\vartheta(x)$ [6], where $J(x)$ is an external
source of the $\vartheta$--field. The collective zero--mode of the
free massless (pseudo)scalar field has been removed by the constraint
[6]
\begin{eqnarray}\label{label1.14}
\int d^2x\,J(x) = \tilde{J}(0) = 0.
\end{eqnarray}
The removal of the collective zero--mode from the intermediate states
of correlation functions, described by the generating functional of
Green functions $Z[J]$, makes it reasonable to delete this mode from
Wightman's observables. For this aim the test functions should obey
the constraint $\tilde{h}(0) = 0$. This is fulfilled for the test
functions $h(x)$ from ${\cal S}_0(\mathbb{R}^{\,2})$. Hence, defining
Wightman's observables on the test functions from ${\cal
S}_0(\mathbb{R}^{\,2}) = \{h(x) \in {\cal S}(\mathbb{R}^{\,2};
\tilde{h}(0) = 0\}$ one makes them insensitive to the collective
zero--mode of the ``center of mass'' of the $\vartheta$--field
described by the Lagrangian (\ref{label1.1}). In other words, the
collective zero--mode of a free massless (pseudo)scalar field cannot
be measured by Wightman's observables.

The interpretation of Wightman's observables suggested in [5] allows
to bridge standard [7,8] and axiomatic quantum field theory
[1--4]. Four years after Wightman's lectures [1], where he declared
that the free massless (pseudo)scalar field $\vartheta(x)$ in
1+1--dimensional space--time does not exist from the point of view of
axiomatic quantum field theory,  Callan, Dashen and Sharp have
published their seminal paper [9] entitled ``Solvable Two--Dimensional
Field Theory Based on Currents''. In this paper the authors
wrote:``Our discussion is not directed toward formal mathematical
questions regarding the model. The results are of interest, instead,
because they provide a tractable model of a theory based on the
currents and the energy--momentum tensor, and because they allow one
to see very readily (a) why the Thirring model is solvable and (b) why
it has trivial physical consequences.''

The last statement concerning the triviality of the massless Thirring
model pointed out first by Wightman [1] and then by Callan, Dashen and
Sharp [9] we have revised recently in [10]. We have shown that the
massless Thirring model possesses a non--trivial non--perturbative
phase of spontaneously broken chiral symmetry. The wave function of
the non--perturbative vacuum in the chirally broken phase is of the
BCS--type [10,11] with fermions acquiring a dynamical mass. Therefore,
the massless Thirring model is by no means trivial and enriched by
non--perturbative phenomena [10].

In spite of Wightman's declaration Callan, Dashen and Sharp analysed
the massless Thirring model, [12] defined by the
Lagrangian\,\footnote{Here $g$ is a dimensionless coupling constant
that can be both positive and negative as well. The field $\psi(x)$ is
a spinor field with two components $\psi_1(x)$ and $\psi_2(x)$. The
$\gamma$--matrices are defined in terms of the well--known $2\times 2$
Pauli matrices $\sigma_1$, $\sigma_2$ and $\sigma_3$: $\gamma^0 =
\sigma_1$, $\gamma^1 = - i\sigma_2$ and $\gamma^5 = \gamma^0\gamma^1 =
\sigma_3$ [10].}
\begin{eqnarray}\label{label1.15}
{\cal L}_{\rm Th}(x) = \bar{\psi}(x)i\gamma^{\mu}\partial_{\mu}\psi(x) -
\frac{1}{2}\,g\,\bar{\psi}(x)\gamma^{\mu}\psi(x)\bar{\psi}(x)
\gamma_{\mu}\psi(x),
\end{eqnarray}
and have expressed the energy--momentum tensor $\theta_{\mu\nu}$ of
the massless Thirring model
\begin{eqnarray}\label{label1.16} 
\theta_{\mu\nu} = \frac{1}{2c}[j_{\mu}(x)j_{\nu}(x) +
j_{\nu}(x)j_{\mu}(x) - g_{\mu\nu}j_{\alpha}(x)j^{\alpha}(x)],
\end{eqnarray}
where $j_{\mu}(x) = \bar{\psi}(x)\gamma_{\mu}\psi(x)$ and $c$ is the
Schwinger term [13], in terms of the massless scalar field $\varphi(x)$
\begin{eqnarray}\label{label1.17}
\theta_{\mu\nu} =
\frac{1}{2}[\partial_{\mu}\varphi(x)\partial_{\nu}\varphi(x) +
\partial_{\nu}\varphi(x)\partial_{\mu}\varphi(x) -
g_{\mu\nu}\partial_{\alpha}\varphi(x)\partial^{\alpha}\varphi(x)]
\end{eqnarray}
with the bosonization rules
\begin{eqnarray}\label{label1.18}
\frac{1}{\sqrt{c}}\,j_0(x^0, x^1) &=& \Pi(x^0,x^1),\nonumber\\
\frac{1}{\sqrt{c}}\,j_1(x^0, x^1) &=& \frac{\partial \varphi(x^0,
x^1)}{\partial x^1}
\end{eqnarray}
solving the current algebra of the massless Thirring model [9] defined
by
\begin{eqnarray}\label{label1.19}
&&[j_0(x^0, x^1), j_0(x^0, y^1)] = 0,\nonumber\\ &&[j_1(x^0, x^1),
j_1(x^0, y^1)] = 0,\nonumber\\ &&[j_0(x^0, x^1), j_1(x^0, y^1)] =
i\,c\,\frac{\partial}{\partial x^1}\delta(x^1 - y^1),
\end{eqnarray}
Using a relation, analogous to that suggested by Morchio, Pierotti and
Strocchi [14]
\begin{eqnarray}\label{label1.20}
\frac{\partial \varphi(x)}{\partial x^{\mu}}
= \varepsilon_{\mu\nu}\frac{\partial \vartheta(x)}{\partial x_{\nu}},
\end{eqnarray}
where $\varepsilon^{\mu\nu}$ is the anti--symmetric tensor defined by
$\varepsilon^{01} = -\varepsilon^{10} = 1$, one can reduce the
bosonization rules (\ref{label1.18}) to a form agreeing with those
suggested in [10]
\begin{eqnarray}\label{label1.21}
\frac{1}{\sqrt{c}}\,j_0(x^0, x^1) &=& \frac{\partial \vartheta(x^0,
x^1)}{\partial x^1},\nonumber\\ \frac{1}{\sqrt{c}}\,j_1(x^0, x^1) &=&
\frac{\partial \vartheta(x^0, x^1)}{\partial x^0} = \Pi(x).
\end{eqnarray}
The Schwinger term $c$ can be expressed in terms of the coupling
constant of the massless Thirring model as follows [10]
\begin{eqnarray}\label{label1.22}
c = \frac{1}{\pi}\,\Big(1 - e^{\textstyle -2\pi/g}\Big).
\end{eqnarray}
This result has been obtained for the chirally broken phase of the
massless Thirring model [10].

Due to the canonical commutation relation 
\begin{eqnarray}\label{label1.23}
[\Pi(x^0,x^1), \varphi(x^0, y^1)] = - [\vartheta(x^0, x^1),
\Pi(x^0,y^1)] = - i\,\delta(x^1 - y^1)
\end{eqnarray}
the spatial derivatives of the $\varphi$ and $\vartheta$ fields
reproduce fully the Schwinger equal--time commutation relation at the
level of the canonical quantum field theory of a free massless
(pseudo)scalar field
\begin{eqnarray}\label{label1.24}
[j_0(x^0,x^1), j_1(x^0, y^1)] &=& \Big[\Pi(x^0,x^1),
\frac{\partial\varphi(x^0, y^1)}{\partial y^1}\Big] =
\Big[\frac{\partial\vartheta(x^0, x^1)}{\partial x^1},
\Pi(x^0,y^1)\Big] =\nonumber\\ &=& i\,c\,\frac{\partial}{\partial
x^1}\delta(x^1 - y^1).
\end{eqnarray}
Without reference to a certain class of test functions, on which
Wightman's positive definiteness condition should be positive and
finite, Callan, Dashen and Sharp comment concerning the
energy--momentum tensor (\ref{label1.17}): ``That is, it is the
energy--momentum tensor of a free massless scalar field. At this
point, one could introduce the Fock representation for the scalar
field, annihilation and creation operators, etc., and verify in detail
that the energy and momentum operators have the expected properties,
but there is little to be gained by going over these well--known
details.''

We would like to emphasize that one of the main open problems of
axiomatic quantum field theory is the construction of a Fock space of
all {\it observable} states of a free massless (pseudo)scalar field in
1+1--dimensional space--time. However, the stress of this problem can
be relaxed if one takes into account that such states can be never
detected due to their sterility and decoupling from everything.

The equivalence between the massless Thirring model and the quantum
field theory of the free massless (pseudo)scalar field, obtained by
Callan, Dashen and Sharp at the level of current algebra, testifies a
one--to--one correspondence between these two theories and the
irrefutable fact that the non--existence of the free massless
(pseudo)scalar field in 1+1--dimensional space--time entails the
non--existence of the massless Thirring model with self--coupled
fermion fields. 

Starting with Klaiber [15] the problem of the solution of the massless
Thirring model was understood as the possibility to evaluate any
correlation function. In his seminal paper [15] Klaiber suggested a
solution of the massless Thirring model in terms of arbitrary
correlation functions. In our recent paper [16] we have given a detail
analysis of Klaiber's operator formalism and Klaiber's solution of the
massless Thirring model. We have displayed weak and strong sides of
Klaiber's results. We have also shown that the infrared cut--off
$\mu$, appearing in the correlation functions in Klaiber's approach,
can be replaced by the ultra--violet cut--off $\Lambda$ by means of a
non--perturbative renormalization of the wave functions of massless
Thirring fermion fields. This evidences that the massless Thirring
model does not suffer from the problem of infrared
divergences. Thereby, in the bosonized version, described by the
quantum field theory of the free massless (pseudo)scalar field, such a
problem should also not exist when it is treated well.

Klaiber's understanding of the solution of the massless Thirring model
was then realized within the path--integral approach [16--18] (see
also [16]) supplemented by the analysis of chiral Jacobians induced by
local chiral rotations [16,19--24]. The path--integral approach is a
nice tool for the evaluation of any correlation function in the
massless Thirring model in terms of degrees of freedom of a free
massless fermion field and two free massless scalar and pseudoscalar
fields [16--18]. The final expressions for correlation functions do
not depend on the infrared cut--off and contain only the ultra--violet
cut--off $\Lambda$. The dependence of correlation functions on the
ultra--violet cut--off $\Lambda$ can be removed and the cut--off
$\Lambda $ can be replaced by a finite arbitrary scale $M$ by means of
a non--perturbative renormalization of the wave functions of massless
Thirring fermion fields, described by the renormalization constant
$Z_2$ [16].

It is important to emphasize that for the solution of the massless
Thirring model within the path--integral approach Wightman's positive
definiteness condition [1--3] and test functions do not concern.

In order to reconcile all of these approaches (i) the axiomatic
quantum field theory based on Wightman's axioms and Wightman's
positive definiteness condition, (ii) the current algebra, using the
equivalence of the massless Thirring model and the Sugawara model,
where currents are dynamical variables, and (iii) the path--integral
we see only one way to assume that for the axiomatic quantum field
theoretic description of the free massless (pseudo)scalar field theory
we should use only test functions from ${\cal S}_0(\mathbb{R}^{\,2})$
as has been pointed out by Wightman [1]. The physical meaning of such
a constraint is the suppression of the collective zero--mode of the
free massless (pseudo)scalar field $\vartheta(x)$ in the definition of
Wightman's observables $\vartheta(h)$. Such a suppression agrees well
with our conclusion that the collective zero--mode of a free massless
(pseudo)scalar field does not affect the dynamics of relative motions
of the system.  Therefore, the definition of Wightman's observables on
the test functions from ${\cal S}_0(\mathbb{R}^{\,2}) = \{h(x) \in
{\cal S}(\mathbb{R}^{\,2}); \tilde{h}(0) = 0\}$ instead of ${\cal
S}(\mathbb{R}^{\,2})$ is well--motivated and does not contradict
Wightman's axioms and Wightman's positive definiteness condition [1].

As has been shown in [6] such a quantum field theory is enriched by
non--perturbative phenomena. Indeed, it possesses a non--trivial phase
of spontaneously broken continuous symmetry (\ref{label1.2})
characterized by non--vanishing spontaneous magnetization.

In this context let us discuss Wightman's observables $\vartheta(h)$
defined on the test functions $h(x) \in {\cal S}(\mathbb{R}^{\,2})$
and ${\cal S}_0(\mathbb{R}^{\,2})$. The generator $Q(x^0)$ responsible
for shifts of the $\vartheta$--field (\ref{label1.2}) is defined by
\begin{eqnarray}\label{label1.25}
Q(x^0) = \int^{\infty}_{-\infty}dx^1\,j_0(x^0,x^1) =
\int^{\infty}_{-\infty}dx^1\,\Pi(x^0,x^1).
\end{eqnarray}
One can show [5] that under the symmetry transformation
(\ref{label1.2}) Wightman's observable (\ref{label1.3}) is changed by
\begin{eqnarray}\label{label1.26}
e^{\textstyle +i\alpha\,Q(x^0)}\,\vartheta(h)\,e^{\textstyle
-i\alpha\,Q(x^0)} = \vartheta(h) + \alpha\int d^2x\,h(x).
\end{eqnarray}
This yields the variation of Wightman's observable
\begin{eqnarray}\label{label1.27}
\delta \vartheta(h) = \alpha\int d^2x\,h(x).
\end{eqnarray}
Thus, for the general case of test functions $h(x) \in {\cal
S}(\mathbb{R}^{\,2})$ Wightman's observable $\vartheta(h)$ is not
invariant under the field--shifts (\ref{label1.2}).

It is important to emphasize that $\delta \vartheta(h)$ given by
(\ref{label1.27}) is not an operator--valued quantity. Therefore, the
vacuum expectation value coincides with the quantity itself
\begin{eqnarray}\label{label1.28}
\langle \Psi_0|\delta \vartheta(h)|\Psi_0\rangle = \delta \vartheta(h)
=\alpha\int d^2x\,h(x).
\end{eqnarray}
Hence, the variation of Wightman's observable $\delta \vartheta(h)$
contains neither the information about spontaneous breaking of
continuous symmetry nor Goldstone bosons.

In order to make this more obvious let us narrow the class of the test
functions from ${\cal S}(\mathbb{R}^{\,2})$ to ${\cal
S}_0(\mathbb{R}^{\,2})$. In this case Wightman's observable
$\vartheta(h)$ becomes invariant under shifts (\ref{label1.2}) and the
variation of Wightman's observable $\delta \vartheta(h)$ is
identically zero, $\delta \vartheta(h) = 0$. However, this does not
give new information about Goldstone bosons and a spontaneously broken
continuous symmetry in addition to that we have got on the class of
the test functions from ${\cal S}(\mathbb{R}^{\,2})$.

In this connection Coleman's theorem [25] asserting the absence of
Goldstone bosons in 1+1--dimensional space-time seems to be
doubtful. Since Coleman has interpreted this theorem too
strong:''$\ldots$ in two dimensions there is no spontaneous breakdown
of continuous symmetries $\ldots$'' [27], we would like to turn to the
analysis of Coleman's theorem in this paper.

The paper is organized as follows. In Section 2 we give a cursory
outline of Wightman's axioms and Wightman's positive definiteness
condition. In Section 3 we consider a free massless (pseudo)scalar
field theory free of infrared divergences and defined on the test
functions from ${\cal S}_0(\mathbb{R}^{\,2})$. In Section 4 we discuss
a canonical quantum field theory of a massless self--coupled
(pseudo)scalar field with current conservation, satisfying Wightman's
axioms and Wightman's positive definiteness condition.  In this
quantum field theory continuous symmetry is spontaneously broken and
Goldstone bosons are quanta of a massless (pseudo)scalar field. In
Section 5 we analyse Coleman's proof and his theorem. In the
Conclusion we summarize the results.

\section{Wightman's axioms and Wightman's positive definiteness 
condition}
\setcounter{equation}{0}

\hspace{0.2in} According to Wightman [1--4]\,\footnote{We cite these
axioms from the textbook by Glimm and Jaffe [4].} any quantum field
theory should satisfy the following set of axioms:
\begin{itemize}
\item {\bf W1} (Covariance). There is a continuous unitary
representation of the imhomogeneous Lorentz group $g \to U(g)$ on the
Hilbert space ${\cal H}$ of quantum theory states. The generators $H =
(P^0, P^1)$ of the translation subgroup have spectrum in the forward
cone $(p^0)^2 - (p^1)^2 \ge 0,\,p^0 \ge 0$. There is a vector
$|\Psi_0\rangle \in {\cal H}$ (the vacuum) invariant under the
operators $U(g)$.

\item {\bf W2} (Observables). There are field operators
$\{\vartheta(h): h(x) \in {\cal S}(\mathbb{R}^{\,2})\}$ densely
defined on ${\cal H}$. The vector $|\Psi_0\rangle$ is in the domain of
any polynomial in the $\vartheta(h)$'s, and the subspace ${\cal H}\,'$
spanned algebraically by the vectors $\{\vartheta(h_1)\ldots
\vartheta(h_n)|\Psi_0\rangle ; n \ge 0, h_i \in {\cal
S}(\mathbb{R}^{\,2})\}$ is dense in ${\cal H}$. The field
$\vartheta(h)$ is covariant under the action of the Lorentz group on
${\cal H}$, and depends linearly on $h$. In particular,
$U^{\dagger}(g) \vartheta(h)U(g) = \vartheta(h_g)$.

\item {\bf W3} (Locality). If the supports of $h(x)$ and
$h\,'(x)$ are space--like separated, then
$[\vartheta(h),\vartheta(h\,')] = 0$ on ${\cal H}\,'$.

\item {\bf W4} (Vacuum). The vacuum vector $|\Psi_0\rangle$ is the
unique vector (up to scalar multiples) in ${\cal H}$ which is
invariant under time translations.
\end{itemize}

These axioms should be supplemented by Wightman's positive
definiteness condition which reads [1--3]
\begin{eqnarray}\label{label2.1}
&&|\!|\Psi|\!|^2 = \Big|\!\Big| \alpha_0|\Psi_0\rangle + \alpha_1\int
d^2x_1\,h(x_1)\,\vartheta(x_1)|\Psi_0\rangle \nonumber\\
&&\hspace{1in} + \frac{\alpha_2}{2!}\int\!\!\!\int
d^2x_1d^2x_2\,h(x_1)
h(x_2)\,\vartheta(x_1)\,\vartheta(x_2)|\Psi_0\rangle +
\ldots\Big|\!\Big|^2 \ge 0
\end{eqnarray}
for all $\alpha_i\in \mathbb{R}^1\,(i=0,1,\ldots)$ and the test
functions $h(x)$ from the the Schwartz class ${\cal
S}(\mathbb{R}^{\,2})$, $h(x) \in {\cal S}(\mathbb{R}^{\,2})$ [1--3],
and $|\Psi_0\rangle$ is a vacuum wave function.

The wave function $|\Psi\rangle$ is a linear superposition of all
quantum states $|\Psi_n\rangle$
\begin{eqnarray}\label{label2.2}
|\Psi_n\rangle = \frac{1}{\sqrt{n!}}\int\!\ldots\!\int d^2x_1\ldots
d^2x_n\,h(x_1)\ldots h(x_n)\,\vartheta(x_1)\ldots
\vartheta(x_n)\,|\Psi_0\rangle,
\end{eqnarray}
which are vectors in the Hilbert space ${\cal H}$ [1--4]. In terms of
the two--point Wightman function the relation (\ref{label2.1}) reads
\begin{eqnarray}\label{label2.3}
\int\!\!\!\int d^2xd^2y\,h^*(x)\,D^{(+)}(x-y)\,h(y) \ge 0,
\end{eqnarray}
which is so called Wightman's positive definiteness condition
[1--3].

\section{A free massless (pseudo)scalar 
field theory without infrared divergences. Path--integral approach}
\setcounter{equation}{0}

\hspace{0.2in} The quantum field theory of the free massless
(pseudo)scalar field $\vartheta(x)$ without infrared divergences has
been developed in Ref.[6] within the path--integral approach. The
removal of infrared divergences is caused by the constraint
(\ref{label1.14}) on external sources $J(x)$ of the
$\vartheta$--field. We have shown that such a theory can be fully
determined by the generating functional of Green functions
\begin{eqnarray}\label{label3.1}
\hspace{-0.3in}Z[J] &=& \Big\langle \Psi_0\Big|{\rm
T}\Big(e^{\textstyle i\int
d^2x\,\vartheta(x)J(x)}\Big)\Big|\Psi_0\Big\rangle =\nonumber\\ &=&
\int {\cal D}\vartheta\,e^{\textstyle i\int
d^2x\,\Big[\frac{1}{2}\,\partial_{\mu}\vartheta(x)\partial^{\mu}
\vartheta(x) + \vartheta(x)J(x)\Big]},
\end{eqnarray}
where ${\rm T}$ is a time--ordering operator.

In terms of $Z[J]$ an arbitrary correlation function of the
$\vartheta$--field can be defined as follows
\begin{eqnarray}\label{label3.2}
&&G(x_1,\ldots,x_n;y_1,\ldots,y_p) = \langle
\Psi_0|F(\vartheta(x_1),\ldots,\vartheta(x_n);\vartheta(y_1),\ldots,
\vartheta(y_p))|\Psi_0\rangle = \nonumber\\ &&=
F\Big(-i\frac{\delta}{\delta J(x_1)},\ldots,-i\frac{\delta}{\delta
J(x_n)};- i\frac{\delta}{\delta J(y_1)},\ldots,-i\frac{\delta}{\delta
J(y_p)}\Big)Z[J]\Big|_{\textstyle J = 0}.
\end{eqnarray}
Relative to the massless Thirring model one encounters the problem of
the evaluation of correlation functions of the following kind
\begin{eqnarray}\label{label3.3}
&&G(x_1,\ldots,x_n;y_1,\ldots,y_p) = \Big\langle \Psi_0\Big|{\rm
T}\Big(\prod^n_{j=1}e^{\textstyle +
i\beta\vartheta(x_j)}\prod^p_{k=1}e^{\textstyle
-i\beta\vartheta(y_k)}\Big)\Big|\Psi_0\Big\rangle=\nonumber\\
&&=\exp\Big\{-i\beta\sum^n_{j = 1}\frac{\delta}{\delta J(x_j)} +
i\beta\sum^p_{k = 1}\frac{\delta}{\delta
J(y_k)}\Big\}Z[J]\Big|_{\textstyle J = 0}.
\end{eqnarray}
Since the path--integral over the $\vartheta$--field (\ref{label3.1})
is Gaussian, it can be evaluated explicitly. The result reads
\begin{eqnarray}\label{label3.4}
Z[J] = \exp\,\Big\{i\,\frac{1}{2}\int
d^2x\,d^2y\,J(x)\,\Delta(x-y; M)\,J(y)\Big\},
\end{eqnarray}
where $\Delta(x-y;M)$, the causal two--point Green function, obeys
the equation
\begin{eqnarray}\label{label3.5}
\Box\,\Delta(x-y;M) = \delta^{(2)}(x-y)
\end{eqnarray}
and relates to the Wightman functions as 
\begin{eqnarray}\label{label3.6}
\Delta(x;M) = i\,\theta(+ x^0)\,D^{(+)}(x; M) +
i\,\theta(-x^0)\,D^{(-)}(x; M),
\end{eqnarray}
where $M$ is a finite scale.

Due to the constraint $\tilde{J}(0) = 0$ (\ref{label1.14}) the
collective zero--mode of the $\vartheta$--field can be deleted from
the intermediate states defining vacuum expectation values
(\ref{label3.2}), therefore the measurement of this configuration in
terms of Wightman's observables $\vartheta(h)$ (\ref{label1.3}),
defined on the class of test functions from ${\cal
S}(\mathbb{R}^{\,2})$, has no physical meaning.

In order to show that in the quantum field theory of a free massless
(pseudo)scalar field without infrared divergences the continuous
symmetry, caused by the field shifts (\ref{label1.2}), is
spontaneously broken we suggest to consider the massless
(pseudo)scalar field $\vartheta(x)$ coupled to an external
``magnetic'' field $h_{\lambda}(x)$ [28], where $h_{\lambda}(x)$ is a
sequence of Schwartz functions from ${\cal S}_0(\mathbb{R}^{\,2})$
with vanishing norm at $\lambda \to \infty$. The Lagrangian
(\ref{label1.1}) should be changed as follows
\begin{eqnarray}\label{label3.7}
{\cal L}(x; h_{\lambda}) =
\frac{1}{2}\,\partial_{\mu}\vartheta(x)\partial^{\mu}\vartheta(x) +
h_{\lambda}(x)\,\vartheta(x).
\end{eqnarray}
The Lagrangian (\ref{label3.7}) defines the action of a massless
(pseudo)scalar field $\vartheta(x)$ coupled to the ``magnetic''field
$h_{\lambda}(x)$
\begin{eqnarray}\label{label3.8}
S[\vartheta, h_{\lambda}] = \int d^2x\,{\cal L}(x; h_{\lambda}) =
\frac{1}{2}\int
d^2x\,\partial_{\mu}\vartheta(x)\partial^{\mu}\vartheta(x) +\int d^2x\,
h_{\lambda}(x)\,\vartheta(x).
\end{eqnarray}
Since the ``magnetic''field $h_{\lambda}(x)$ belongs to the Schwartz 
class ${\cal S}_0(\mathbb{R}^{\,2})$ obeying the constraint
\begin{eqnarray}\label{label3.9}
\int d^2x\,h_{\lambda}(x) = \tilde{h}_{\lambda}(0) = 0,
\end{eqnarray}
the action $S[\vartheta, h_{\lambda}]$ is invariant under the symmetry
transformation (\ref{label1.2}).

Making a field--shift (\ref{label1.2}) we get
\begin{eqnarray}\label{label3.10}
\hspace{-0.3in}&&S[\vartheta, h_{\lambda}] \to S\,'[\vartheta,
h_{\lambda}] = \frac{1}{2}\int
d^2x\,\partial_{\mu}\vartheta\,'(x)\partial^{\mu}\vartheta\,'(x) +\int
d^2x\, h_{\lambda}(x)\,\vartheta\,'(x) =\nonumber\\ \hspace{-0.3in}&&=
S[\vartheta, h_{\lambda}] + \alpha\int d^2x\,h_{\lambda}(x).
\end{eqnarray}
Due to the constraint (\ref{label3.9}) the r.h.s. of (\ref{label3.10})
is equal to $S[\vartheta, h_{\lambda}]$. This confirms the invariance
of the action under the symmetry transformations (\ref{label1.2}).

The generating functional of Green functions reads now
\begin{eqnarray}\label{label3.11}
\hspace{-0.3in}Z[J;h_{\lambda}] &=& \Big\langle \Psi_0\Big|{\rm
T}\Big(\,e^{\textstyle i\int d^2x\,\vartheta(x)(h_{\lambda}(x) +
J(x))}\Big)\Big|\Psi_0\Big\rangle =\nonumber\\ 
\hspace{-0.3in}&=&\int {\cal D}\vartheta\,e^{\textstyle i\int
d^2x\,\Big[\frac{1}{2}\,\partial_{\mu}\vartheta(x)\partial^{\mu}
\vartheta(x) + \vartheta(x)(h_{\lambda}(x) + J(x))\Big]} = \nonumber\\
\hspace{-0.3in}&=& \exp\,\Big\{i\,\frac{1}{2}\int
d^2x\,d^2y\,(h_{\lambda}(x) + J(x))\,\Delta(x-y; M)\,(h_{\lambda}(y) +
J(y))\Big\}.
\end{eqnarray}
We remind that due the constraints $\tilde{J}(0) =
\tilde{h}_{\lambda}(0) = 0$ the generating functional of Green
functions $Z[J;h_{\lambda}]$ is invariant under field--shifts (\ref{label1.2}).

According to Itzykson and Drouffe [28] the magnetization ${\cal
M}(h_{\lambda})$ can be defined by [6]\,\footnote{For simplicity we
consider symmetric functions $\tilde{h}_{\lambda}(k_+,k_-) =
\tilde{h}_{\lambda}(k_-,k_+)$.}
\begin{eqnarray}\label{label3.12}
&&{\cal M}(h_{\lambda}) = \langle
\Psi_0|\cos\vartheta(h_{\lambda})|\Psi_0\rangle = \exp \Big\{ -
\frac{1}{2}\int d^2x\,d^2y\,h^*_{\lambda}(x)\,D^{(+)}(x-y;
M)\,h_{\lambda}(y)\Big\} = \nonumber\\ &&= \exp \Big\{ -
\frac{1}{4\pi}\int^{\infty}_0\frac{dk_+}{k_+}\,
|\tilde{h}_{\lambda}(k_+, 0)|^2\,\Big\}.
\end{eqnarray}
Since the operator $\cos\vartheta(h_{\lambda})$ is not time--ordered,
the vacuum expectation value (\ref{label3.12}) is defined in terms of
the Wightman function $D^{(+)}(x-y; M)$ (\ref{label1.7}), with the
replacement of the infrared cut--off $\mu$ by the finite scale $M$
[6], but not the causal Green function $\Delta(x-y; M)$ of
(\ref{label3.6}).

We would like to emphasize that the exponent in the r.h.s. of
(\ref{label3.12}) has the form of Wightman's positive definiteness
condition (\ref{label1.10}). Due to the fast decreasing of the
functions $\tilde{h}_{\lambda}(k_+,0)$ for $k_+ \to \infty$ and the
constraint $\tilde{h}_{\lambda}(0,0) = 0$, the integral over $k_+$ is
convergent and positive definite.

If we switch off the ``magnetic'' field taking the limit $h_{\lambda}
\to 0$, this can be done adiabatically defining $h_{\lambda}(x) =
e^{\textstyle -\varepsilon\,\lambda}\,h(x)$ for $\lambda \to \infty$
with $\varepsilon$, a positive, infinitesimally small parameter, we
get
\begin{eqnarray}\label{label3.13}
{\cal M} = \lim_{\lambda \to \infty}{\cal M}(h_{\lambda}) = 1.
\end{eqnarray}
This agrees with our results obtained in [6]. The quantity ${\cal M}$
is the spontaneous magnetization. Since the spontaneous magnetization
does not vanish, ${\cal M} = 1$, the continuous symmetry, caused by the
field--shifts (\ref{label1.2}), is spontaneously broken. This confirms
our statement concerning the existence of the chirally broken phase in
the massless Thirring model [10].

Thus, we have shown that in the quantum field theory of a free
massless (pseudo)scalar field $\vartheta(x)$, defined on test
functions $h(x)$ from ${\cal S}_0(\mathbb{R}^{\,2})$ and external
sources $J(x)$, obeying the constraint $\tilde{J}(0) = 0$, there is a
non--perturbative phase, characterized by a non--vanishing spontaneous
magnetization ${\cal M} = 1$, testifying the existence of a
spontaneously broken continuous symmetry (\ref{label1.2}). 

\section{Canonical quantum field theory of a massless 
self--coupled (pseudo)scalar field} 
\setcounter{equation}{0}

\hspace{0.2in} For the most general version of a canonical quantum
field theory, which we consider as a candidate for a test of Coleman's
theorem, we assume a quantum field theory of a massless self--coupled
(pseudo)scalar field $\vartheta(x)$ with current conservation
$\partial_{\mu}j^{\mu}(x) = 0$ satisfying Wightman's axioms {\bf W1}
-- {\bf W4} and Wightman's positive definiteness condition on the test
functions $h(x)$ from ${\cal S}(\mathbb{R}^{\,2})$ [5]. Nevertheless,
below we will deal only with test functions from ${\cal
S}_0(\mathbb{R}^{\,2}) = \{h(x) \in {\cal S}(\mathbb{R}^{\,2});
\tilde{h}(0) = 0\}$. The most useful tool for the analysis of this
theory is the K\"allen--Lehmann representation [29].

For the proof of his theorem Coleman has taken the quantum field
theory of a massless (pseudo)scalar field $\vartheta(x)$ with a
conserved current $\partial^{\mu}j_{\mu}(x) = 0$ and considered the
Fourier transforms $F^{(+)}(k)$, $F^{(+)}_{\mu}(k)$ and
$F^{(+)}_{\mu\nu}(k)$ of the two--point functions defined by
\begin{eqnarray}\label{label4.1}
F^{(+)}(k) &=&\int d^2x\,e^{\textstyle i\,k\cdot x}\,D^{(+)}(x)
= \int d^2x\,e^{\textstyle i\,k\cdot x}\,\langle \Psi_0
|\vartheta(x)\vartheta(0)|\Psi_0 \rangle,\nonumber\\ F^{(+)}_{\mu}(k)
&=&i\int d^2x\,e^{\textstyle i\,k\cdot x}\,D^{(+)}_{\mu}(x) =
i\int d^2x\,e^{\textstyle i\,k\cdot x}\,\langle \Psi_0
|j_{\mu}(x)\vartheta(0)|\Psi_0 \rangle,\nonumber\\ F^{(+)}_{\mu\nu}(k)
&=&\int d^2x\,e^{\textstyle i\,k\cdot x}\,D^{(+)}_{\mu\nu}(x) =
\int d^2x\,e^{\textstyle i\,k\cdot x}\,\langle \Psi_0
|j_{\mu}(x)j_{\nu}(0)|\Psi_0 \rangle,
\end{eqnarray}
where $D^{(+)}(x)$, $D^{(+)}_{\mu}(x)$ and
$D^{(+)}_{\mu\nu}(x)$ are the Wightman function
(\ref{label1.7}), current--field and current--current correlation
functions calculated with respect to the vacuum state
$|\Psi_0\rangle$, invariant under space and time translations (see
Wightman's axioms).

Since only the Fourier transform $F^{(+)}_{\mu}(k)$ can test Goldstone
bosons we turn to the consideration of this function only. Inserting a
complete set of intermediate states, the Fourier transform
$F^{(+)}_{\mu}(k)$ can be transcribed into the form
\begin{eqnarray}\label{label4.2}
F^{(+)}_{\mu}(k) = i\sum_{n}\int d^2x\,e^{\textstyle +
ik\cdot x}\, \langle \Psi_0|j_{\mu}(x)|n\rangle\langle
n|\vartheta(0)|\Psi_0\rangle,
\end{eqnarray}
Due to the invariance of the vacuum state $|\Psi_0\rangle$ under space
and time translations and Lorentz covariance
$\langle\Psi_0|j_{\mu}(x)|\Psi_0\rangle = 0$, we have
\begin{eqnarray}\label{label4.3}
F^{(+)}_{\mu}(k) &=& i\sum_{n\neq \Psi_0} \int
d^2x\,e^{\textstyle + ik\cdot x}\, \langle
\Psi_0|j_{\mu}(x)|n\rangle\langle
n|\vartheta(0)|\Psi_0\rangle.
\end{eqnarray}
Using again the invariance of the vacuum state $|\Psi_0\rangle$ under
space and time translations we obtain [29]
\begin{eqnarray}\label{label4.4}
F^{(+)}_{\mu}(k)
&=&i(2\pi)^2 \sum_{n\neq \Psi_0 }\delta^{(2)}(k -
p_n)\langle\Psi_0|j_{\mu}(0)|n\rangle\langle n|\vartheta(0)|
\Psi_0\rangle
\end{eqnarray}
The r.h.s. can be rewritten in the form of the K\"allen--Lehmann
representation in terms of the spectral function $\rho(m^2)$ which is
defined by [29]
\begin{eqnarray}\label{label4.5}
\hspace{-0.3in}F^{(+)}_{\mu}(k)&=&i\,(2\pi)^2\sum_{n\neq
\Psi_0}\delta^{(2)}(k - p_n)\langle\Psi_0|j_{\mu}(0)|n\rangle\langle
n|\vartheta(0)|\Psi_0\rangle = \nonumber\\
\hspace{-0.3in}&=& - \varepsilon_{\mu\nu}\,k^{\nu}\,
\varepsilon(k^1)\,\theta(k^0) \int^{\infty}_0\!\!\!\delta(k^2 -
m^2)\,\rho(m^2)dm^2.
\end{eqnarray}
We notice that for massless states $-
\varepsilon_{\mu\nu} \,k^{\nu}\, \varepsilon(k^1) = k_{\mu}$.

This is the most general form of a {\it tempered} distribution in the
quantum field theory of a massless self--coupled (pseudo)scalar field
$\vartheta(x)$ with current conservation $\partial^{\mu}j_{\mu}(x) =
0$ in 1+1--dimensional space--time satisfying Wightman's axioms {\bf
W1} -- {\bf W4} and Wightman's positive definiteness condition on the
test functions $h(x)$ from ${\cal S}(\mathbb{R}^{\,2})$.

Let us isolate the contribution of the state with $m^2 = 0$ to
$F^{(+)}_{\mu}(k)$.  Setting the spectral function $\rho(m^2)$ equal
to
\begin{eqnarray}\label{label4.6}
\rho(m^2) &=& \sigma\,\delta(m^2) + \rho\,'(m^2),
\end{eqnarray}
we obtain the Fourier transform $F^{(+)}_{\mu}(k)$ in the form
\begin{eqnarray}\label{label4.7}
\hspace{-0.3in}F^{(+)}_{\mu}(k)
&=&\sigma k_{\mu} \theta(k^0) \delta(k^2) -
\varepsilon_{\mu\nu}\,k^{\nu}\,\varepsilon(k^1)\,
\theta(k^0)\int^{\infty}_{M^2} \delta(k^2 -
m^2) \rho\,'(m^2)dm^2,
\end{eqnarray}
where the spectral function $\rho\,'(m^2)$
contains only the contributions of the states with $m^2 > 0$ and the
scale $M^2$ separates the state with $m^2 = 0$ from the states with
$m^2 > 0$.

The original of the Fourier transform given by (\ref{label4.5}) is
given by
\begin{eqnarray}\label{label4.8}
iD^{(+)}_{\mu}(x) = -i
\varepsilon_{\mu\nu}\frac{\partial}{\partial
x_{\nu}}\int^{\infty}_0\frac{dm^2}{8\pi^2}\,\rho(m^2)
\int^{\varphi_0}_{-\varphi_0}d\varphi\,e^{\textstyle -m\sqrt{-x^2 +
i0\cdot \varepsilon(x^0)}\,{\cosh}\varphi},
\end{eqnarray}
where $\varphi_0$ is defined by [16] 
\begin{eqnarray}\label{label4.9}
\varphi_0 = \frac{1}{2}\,{\ell n}\Big(\frac{x^0 + x^1 - i0}{x^0 - x^1
- i0}\Big).
\end{eqnarray}
No we transcribe the r.h.s. of $iD^{(+)}_{\mu}(x)$ as follows
\begin{eqnarray}\label{label4.10}
\hspace{-0.3in}&&iD^{(+)}_{\mu}(x) = -i
\varepsilon_{\mu\nu}\frac{\partial \varphi_0}{\partial
x_{\nu}}\Big[\frac{1}{4\pi^2}\int^{\infty}_0dm^2\,\rho(m^2)\Big]
\nonumber\\ \hspace{-0.3in}&& - i
\varepsilon_{\mu\nu}\frac{\partial}{\partial
x_{\nu}}\frac{1}{8\pi^2}\int^{\infty}_0dm^2\,\rho(m^2)
\int^{\varphi_0}_{-\varphi_0}d\varphi\,\Big(e^{\textstyle -m\sqrt{-x^2
+ i0\cdot \varepsilon(x^0)}\,{\cosh}\varphi} - 1\Big)=\nonumber\\
\hspace{-0.3in}&&= \Big[\frac{1}{2\pi}\int^{\infty}_0
dm^2\,\rho(m^2)\Big] \frac{i}{2\pi}\,\frac{x_{\mu}}{-x^2 + i0\cdot
\varepsilon(x^0)} \nonumber\\ \hspace{-0.3in}&& - i
\varepsilon_{\mu\nu}\frac{\partial}{\partial
x_{\nu}}\frac{1}{8\pi^2}\int^{\infty}_0dm^2\,\rho(m^2)
\int^{\varphi_0}_{-\varphi_0}d\varphi\,\Big(e^{\textstyle -m\sqrt{-x^2
+ i0\cdot \varepsilon(x^0)}\,{\cosh}\varphi} - 1\Big)=\nonumber\\
\hspace{-0.3in}&&= \int^{\infty}_0dm^2\,\rho(m^2)\int
\frac{d^2q}{(2\pi)^2}\,q_{\mu}\,\theta(q^0)\,\delta(q^2)\,e^{\textstyle
-iq\cdot x}\nonumber\\ \hspace{-0.3in}&& - i
\varepsilon_{\mu\nu}\frac{\partial}{\partial
x_{\nu}}\frac{1}{8\pi^2}\int^{\infty}_0dm^2\,\rho(m^2)
\int^{\varphi_0}_{-\varphi_0}d\varphi\,\Big(e^{\textstyle -m\sqrt{-x^2
+ i0\cdot \varepsilon(x^0)}\,{\cosh}\varphi} - 1\Big).
\end{eqnarray}
This defines $iD^{(+)}_{\mu}(x)$ in the following general form
\begin{eqnarray}\label{label4.11}
\hspace{-0.3in}&&iD^{(+)}_{\mu}(x)
=\int^{\infty}_0dm^2\,\rho(m^2)\int
\frac{d^2q}{(2\pi)^2}\,q_{\mu}\,\theta(q^0)\,\delta(q^2)\,e^{\textstyle
-iq\cdot x}\nonumber\\ \hspace{-0.3in}&& - i
\varepsilon_{\mu\nu}\frac{\partial}{\partial
x_{\nu}}\frac{1}{8\pi^2}\int^{\infty}_0dm^2\,\rho(m^2)
\int^{\varphi_0}_{-\varphi_0}d\varphi\,\Big(e^{\textstyle -m\sqrt{-x^2
+ i0\cdot \varepsilon(x^0)}\,{\cosh}\varphi} - 1\Big).
\end{eqnarray}
The first term describes the contribution of the state with $m^2 = 0$,
whereas the second one contains the contributions of all states with
$m^2 > 0$.  Since the contribution of the state with $m^2 = 0$ is
defined by the expression $F^{(+)}_{\mu}(k; m^2 = 0)
=\sigma\,k_{\mu}\,\theta(k^0)\,\delta(k^2)$ [5,25], we get sum rules
for the spectral function $\rho(m^2)$, which read
\begin{eqnarray}\label{label4.12}
\int^{\infty}_0dm^2\,\rho(m^2) = \sigma.
\end{eqnarray}
In order to investigate further properties of the spectral function
$\rho(m^2)$ we suggest to consider the vacuum expectation value
$\langle \Psi_0|[j_{\mu}(x), \vartheta(0)]| \Psi_0 \rangle$. Following
the standard procedure expounded above we get
\begin{eqnarray}\label{label4.13}
\hspace{-0.5in}&&\langle
\Psi_0|[j_{\mu}(x),\vartheta(0)]|\Psi_0\rangle = \nonumber\\
\hspace{-0.5in}&& = i\int^{\infty}_0dm^2\,\rho(m^2)\int
\frac{d^2k}{(2\pi)^2}\,\varepsilon_{\mu\nu}\,k^{\nu}\,
\varepsilon(k^1)\,\theta(k^0)\,\delta(k^2 - m^2)\,(e^{\textstyle
-ik\cdot x} + e^{\textstyle +ik\cdot x}).
\end{eqnarray}
The vacuum expectation value of the equal--time commutation relation
for the time--component of the current $j_0(0,x^1)$ and the field
$\vartheta(0)$ reads
\begin{eqnarray}\label{label4.14}
&&\langle \Psi_0|[j_0(0,x^1),\vartheta(0)]|\Psi_0\rangle =
-\frac{1}{2\pi}\int^{\infty}_0dm^2\,\rho(m^2)\,i\,\delta(x^1)
\nonumber\\&&- i\int^{\infty}_0dm^2\,\rho(m^2)\int^{\infty}_0
\frac{dk^1}{2\pi^2}\Big(\frac{k^1}{\sqrt{(k^1)^2 + m^2}} -
1\Big)\,\cos(k^1x^1).
\end{eqnarray}
The state with $m^2 = 0$ does not contribute to the second term. This
term is defined by the spectral function $\rho\,'(m^2)$ only 
\begin{eqnarray}\label{label4.15}
&&\langle \Psi_0|[j_0(0,x^1),\vartheta(0)]|\Psi_0\rangle =
-\frac{1}{2\pi}\int^{\infty}_0dm^2\,\rho(m^2)\,i\,\delta(x^1)
\nonumber\\&&- i\int^{\infty}_{M^2}dm^2\,\rho\,'(m^2)\int^{\infty}_0
\frac{dk^1}{2\pi^2}\Big(\frac{k^1}{\sqrt{(k^1)^2 + m^2}} -
1\Big)\,\cos(k^1x^1).
\end{eqnarray}
Let us show that in the canonical quantum field theory the second term
in (\ref{label4.15}) should vanish.

For this aim it is sufficient to prove that in the quantum field
theory of a massless (pseudo)scalar field $\vartheta(x)$ the canonical
conjugate momentum $\Pi(x)$ coincides with the time component of the
current $j_{\mu}(x)$, i.e. $\Pi(x) = j_0(x)$. This results in the
l.h.s. of (\ref{label4.15}) equal to $-i\,\delta(x^1)$.

The fact that $j_0(x)$ is equal to the conjugate momentum $\Pi(x)$ of
the field $\vartheta(x)$ can be easily illustrated in terms of the
Lagrangian. The general Lagrangian of the massless self--coupled
(pseudo)scalar field $\vartheta(x)$, invariant under the field--shifts
$\vartheta(x) \to \vartheta\,'(x) = \vartheta(x) + \alpha$, should
depend only on $\partial_{\mu}\vartheta(x)$ and can be written as
\begin{eqnarray}\label{label4.16}
{\cal L}(x) = {\cal L}[\partial_{\mu}\vartheta(x)].
\end{eqnarray}
In the Lagrange approach the current $j_{\mu}(x)$ is defined by 
\begin{eqnarray}\label{label4.17}
j_{\mu}(x) = \frac{\delta {\cal L}[\partial_{\mu}\vartheta(x)]}{\delta
\partial^{\mu}\vartheta(x)}.
\end{eqnarray}
This testifies the coincidence of $j_0(x)$ with the conjugate momentum
$\Pi(x)$, which is the derivative of the Lagrangian with respect to
the time--derivative of the $\vartheta$--field,
$\dot{\vartheta}(x)$. We get
\begin{eqnarray}\label{label4.18}
\Pi(x)= \frac{\delta {\cal L}[\partial_{\mu}\vartheta(x)]}{\delta
\dot{\vartheta}(x)} = j_0(x).
\end{eqnarray}
Using the canonical equal--time commutation relation 
\begin{eqnarray}\label{label4.19}
[j_0(0,x^1),\vartheta(0)] = [\Pi(0,x^1),\vartheta(0)] = -i\delta(x^1)
\end{eqnarray}
we derive from (\ref{label4.15}) the sum rules
\begin{eqnarray}\label{label4.20}
\int^{\infty}_0dm^2\,\rho(m^2) = 2\pi.
\end{eqnarray}
Comparing (\ref{label4.20}) with (\ref{label4.12}) we get
\begin{eqnarray}\label{label4.21}
\sigma = 2\pi.
\end{eqnarray}
This rules out Coleman's result asserting $\sigma = 0$.

As the second term in the r.h.s. of (\ref{label4.15}) can be never 
proportional to $\delta(x^1)$ it should be zero. This yields 
\begin{eqnarray}\label{label4.22}
\rho\,'(m^2) \equiv 0.
\end{eqnarray}
Hence, the spectral function $\rho(m^2)$ is equal to
\begin{eqnarray}\label{label4.23}
\rho(m^2) = \sigma\,\delta(m^2) = 2\pi\,\delta(m^2).
\end{eqnarray}
This means that in the case of current conservation
$\partial^{\mu}j_{\mu}(x) = 0$ the Fourier transform
$F^{(+)}_{\mu}(k)$ is defined by the contribution of the state with
$m^2 = 0$ only. This confirms that the expression $F^{(+)}_{\mu}(k) =
\sigma\,k_{\mu}\,\theta(k^0)\,\delta(k^2)$, postulated by Coleman
[25], is general for canonical quantum field theories with conserved
current $\partial^{\mu}j_{\mu}(x) = 0$ but rules out Coleman's result
$\sigma = 0$ [5]. In non--canonical quantum field theories the
expression $F^{(+)}_{\mu}(k) = \sigma\, k_{\mu}\, \theta(k^0)\,
\delta(k^2)$, postulated by Coleman, is not general and should be
rewritten in the form (\ref{label4.7}) with $\rho\,'(m^2) \neq
0$. Hence, Coleman's result can only be understood by the fact that he
removed the canonical massless field $\vartheta(x)$ from the
consideration.

Multiplying (\ref{label4.15}) by $i\,\alpha$ and integrating over
$x^1$ we obtain $\langle \Psi_0|\delta \vartheta(0)|\Psi_0 \rangle$,
which reads
\begin{eqnarray}\label{label4.24}
\langle \Psi_0|\delta \vartheta(0)|\Psi_0 \rangle =
i\,\alpha\int^{\infty}_{-\infty}dx^1\,\langle
\Psi_0|[j_0(0,x^1),\vartheta(0)]|\Psi_0\rangle = \frac{\alpha}{2\pi}
\int^{\infty}_0dm^2\,\rho(m^2) = \alpha,
\end{eqnarray}
where we have taken into account the discussion above and the
expression for the spectral function $\rho(m^2)$ given by
(\ref{label4.23}).

\section{Triviality of Coleman's proof of his theorem}
\setcounter{equation}{0}

\hspace{0.2in} In this section we would like to show that Coleman's
theorem is a trivial consequence of the exclusions of massless
one--particle states and has no relation to the suppression of
spontaneous symmetry breakdown. 

For the proof of his theorem Coleman treated the Cauchy--Schwarz
inequality
\begin{eqnarray}\label{label5.1}
\int \frac{d^2k}{2\pi}\,|\tilde{h}_{\lambda}(k)|^2\,F^{(+)}(k)\int
\frac{d^2k}{2\pi}\,|\tilde{h}_{\lambda}(k)|^2\,F^{(+)}_{00}(k)\ge
\Big[\int
\frac{d^2k}{2\pi}\,|\tilde{h}_{\lambda}(k)|^2\,F^{(+)}_0(k)\Big]^2,
\end{eqnarray}
defined on the test functions $h_{\lambda}(x)$ from ${\cal
S}(\mathbb{R}^{\,1})\otimes {\cal S}_0(\mathbb{R}^{\,1})$:
\begin{eqnarray}\label{label5.2}
h_{\lambda}(x) =
\frac{1}{\lambda}\,f\Big(\frac{x_+}{\lambda}\Big)\,g(x_-) + 
\frac{1}{\lambda}\,f\Big(\frac{x_-}{\lambda}\Big)\,g(x_+)
\end{eqnarray}
the Fourier transform of which is given by
\begin{eqnarray}\label{label5.3}
\tilde{h}_{\lambda}(k_+,k_-) = \tilde{f}(\lambda k_-)\,\tilde{g}(k_+) +
\tilde{f}(\lambda k_+)\,\tilde{g}(k_-)
\end{eqnarray}
with $f(x_{\pm}) \in {\cal S}(\mathbb{R}^{\,1})$ and $g(x_{\pm}) \in
{\cal S}_0(\mathbb{R}^{\,1})$.

In order to make Coleman's exclusions more transparent we suggest to
rewrite the inequality (\ref{label5.1}) in the equivalent form
\begin{eqnarray}\label{label5.4}
&&\int d^2xd^2y\,h^*(x)\,D^{(+)}(x-y)\,h(y)\int
d^2xd^2y\,h^*(x)\,D^{(+)}_{00}(x-y)\,h(y)\nonumber\\ && \ge \Big[\int
d^2xd^2y\,h^*(x)\,iD^{(+)}_0(x-y)\,h(y)\Big]^2.
\end{eqnarray}
In terms of vacuum expectation values it reads
\begin{eqnarray}\label{label5.5}
&&\Big[\int d^2xd^2y\,h^*(x)\,\langle
\Psi_0|\vartheta(x)\vartheta(y)|\Psi_0\rangle\,h(y)\Big]\Big[\int
d^2xd^2y\,h^*(x)\,\langle
\Psi_0|j_0(x)j_0(y)|\Psi_0\rangle\,h(y)\Big]\nonumber\\ && \ge
\Big[\int d^2xd^2y\,h^*(x)\,i\langle
\Psi_0|j_0(x)\vartheta(y)|\Psi_0\rangle\,h(y)\Big]^2.
\end{eqnarray}
Inserting a complete set of intermediate states, the eigenstates
$|n\rangle$ of the full Hamiltonian, we get
\begin{eqnarray}\label{label5.6}
&&\sum_n\int d^2xd^2y\,h^*(x)\,\langle
\Psi_0|\vartheta(x)|n\rangle\langle
n|\vartheta(y)|\Psi_0\rangle\,h(y)\nonumber\\ &&\times\sum_n\int
d^2xd^2y\,h^*(x)\,\langle \Psi_0|j_0(x)|n\rangle\langle
n|j_0(y)|\Psi_0\rangle\,h(y)\nonumber\\ && \ge \Big[\sum_n\int
d^2xd^2y\,h^*(x)\,i\langle \Psi_0|j_0(x)|n\rangle\langle
n|\vartheta(y)|\Psi_0\rangle\,h(y)\Big]^2.
\end{eqnarray}
Using the invariance of the vacuum state under space and time
translations and Lorentz covariance we get
\begin{eqnarray}\label{label5.7}
&&\sum_n |\tilde{h}(p_n)|^2 |\langle
n|\vartheta(0)|\Psi_0\rangle|^2\sum_n|\tilde{h}(p_n)|^2|\langle
n|j_0(0)|\Psi_0\rangle|^2\nonumber\\ && \ge \Big[\sum_n
|\tilde{h}(p_n)|^2i\langle \Psi_0|j_0(0)|n\rangle\langle
n|\vartheta(0)|\Psi_0\rangle\Big]^2.
\end{eqnarray}
Now it is convenient to introduce the following notations
\begin{eqnarray}\label{label5.8}
&&\sum_n |\tilde{h}(p_n)|^2 \,|\langle n|\vartheta(0)|\Psi_0\rangle|^2
= \langle \Psi_0|\vartheta(0)|\Psi_0\rangle^2|\tilde{h}(0)|^2
\nonumber\\&& + \sum_{n\neq \Psi_0, p^2_n = 0} |\tilde{h}(p_n)|^2
|\langle n|\vartheta(0)|\Psi_0\rangle|^2 + \sum_{n\neq \Psi_0, p^2_n
\neq 0} |\tilde{h}(p_n)|^2 |\langle n|\vartheta(0)|\Psi_0\rangle|^2
=\nonumber\\ &&= a^2_0 + a^2_1 + a^2_2,\nonumber\\
&&\sum_n |\tilde{h}(p_n)|^2 \,|\langle
n|j_0(0)|\Psi_0\rangle|^2 = \nonumber\\ &&= \sum_{ n\neq \Psi_0, p^2_n
= 0} |\tilde{h}(p_n)|^2 |\langle n|j_0(0)|\Psi_0\rangle|^2 +
\sum_{n\neq \Psi_0, p^2_n \neq 0} |\tilde{h}(p_n)|^2 |\langle
n|j_0(0)|\Psi_0\rangle|^2=\nonumber\\
&&\vspace{0.1in}=b^2_1 + b^2_2,\nonumber\\ &&\sum_n
|\tilde{h}(p_n)|^2 \,i\,\langle \Psi_0|j_0(0)|n\rangle\langle
n|\vartheta(0)|\Psi_0\rangle = \nonumber\\ 
&&=\sum_{n\neq \Psi_0, p^2_n = 0}
|\tilde{h}(p_n)|^2\,i\,\langle \Psi_0|j_0(0)|n\rangle \langle
n|\vartheta(0)|\Psi_0 \rangle\nonumber\\ && + \sum_{n\neq \Psi_0,
p^2_n \neq 0} |\tilde{h}(p_n)|^2\,i\,\langle \Psi_0|j_0(0)|n\rangle
\langle n|\vartheta(0)|\Psi_0 \rangle = a_1b_1 + a_2b_2,
\end{eqnarray}
where the indices $i=0,1,2$ correspond to the
contributions of the vacuum state, the massless state with $p^2_n = 0$
and the states with $p^2_n \neq 0$, respectively.

In terms of the vectors $\vec{a} =
(a_0, a_1, a_2)$ and $\vec{b} = (0, b_1, b_2)$ the inequality
(\ref{label5.7}) reads
\begin{eqnarray}\label{label5.9}
(a^2_0 + a^2_1 + a^2_2)(b^2_1 + b^2_2)\ge (a_1b_1 +
a_2b_2)^2.
\end{eqnarray}
This is the Cauchy--Schwarz inequality for $\vec{a}$ and $\vec{b}$.

The inequality (\ref{label5.9}) is still correct if an arbitrary
number of intermediate states is removed from the sums $\sum_n$. In
his proof Coleman has used this fact carrying out the following steps:

\begin{itemize}
\item (i) $a_1 = 0$: Due to the extension of test functions from
${\cal S}_0(\mathbb{R}^{\,2})$ to ${\cal S}(\mathbb{R}^{\,2})$
zero--mass contributions to the Wightman functions are excluded from
the very beginning.
\item (ii) $a_0 = 0$: By the special choice of the test function $g(0)
= 0$.
\item (iii) $b_1 = 0$: By claiming that ``Because $F^{(+)}_{00}(k)$ is
a positive distribution, the second integral is monotone decreasing''
for $\lambda$ going to infinity. Coleman refers with ``the second
integral'' to the expression $\int
d^2k\,|\tilde{h}_{\lambda}(k)|^2F_{00}(k)$.
\item (iv) $a_2 = b_2 = 0$: By the limit $\lambda \to \infty$
removing all intermediate states with non--zero squared invariant
masses, $p^2_n\neq 0$. 
\end{itemize}

Thus, in the limit $\lambda \to \infty$ the Cauchy--Schwarz inequality
(\ref{label5.9}) reads
\begin{eqnarray}\label{label5.10}
0\cdot0 \ge (0\cdot 0)^2 .
\end{eqnarray}
Hence, Coleman removed step by step all intermediate states. Since no
eigenstates of the Hamiltonian are left, there are no contributions to
the r.h.s. of (\ref{label5.1}) and (\ref{label5.10}) and consequently
no massless bosons.  But this does not say anything what happens if the
zero--mass modes are not removed.

The trivial conclusion of these steps is: {\it If one excludes
zero--mass modes from the spectrum of the full Hamiltonian, i.e. from
the intermediate states, they are not present in the theory.} In fact,
since we have shown that in a canonical quantum field theory the only
contributions to $F^{(+)}_0(k)$ come from zero--mass modes, an
exclusion of these modes leads to $F^{(+)}_0(k) = 0$.

Thus, we argue that if one understands Coleman's paper as a proof of
the absence of Goldstone bosons in two dimensions than this conclusion
is wrong.  Goldstone bosons in Coleman's paper are excluded by
Wightman's positive definiteness condition, which he requests for test
functions from ${\cal S}(\mathbb{R}^{\,2})$.

\section{Conclusion}
\setcounter{equation}{0}

\hspace{0.2in} We argue that the inconsistencies of Coleman's theorem
are shortly in the following:
\begin{itemize}
\item Coleman's theorem contradicts canonical quantum field theory of
a massless self--coupled (pseudo)scalar field $\vartheta(x)$ with
current conservation $\partial^{\mu}j_{\mu}(x) = 0$ in which the
parameter $\sigma = 2\pi$ but not $\sigma = 0$. As has been shown in
[5] such a quantum field theory satisfies Wightman's axioms and
Wightman's positive definiteness condition on the test functions
$h(x)$ from ${\cal S}(\mathbb{R}^{\,2})$.

\item Coleman's theorem testifies the obvious and trivial assertion:
``If one removes a canonical massless (pseudo)scalar field from the
theory, this field does not appear in the further consideration.''
\end{itemize}

Then, accepting Coleman's theorem as the proof of the absence of the
quantum field theory of a free massless (pseudo)scalar field, one can
be confused by the following consequence of this theorem demanding the
absence of the quantum field theory of Thirring fermion fields.  In
fact, since the massless Thirring model bosonizes to the quantum field
theory of a free massless (pseudo)scalar field, any suppression of
this quantum field theory would lead to the suppression of the
massless Thirring model. But in this case how do we have to understand
the results obtained within current algebra and path--integral
approach?

Therefore, the only way to reconcile different approaches to the
description of the quantum field theory of a free massless
(pseudo)scalar field defined in 1+1--dimensional space--time: (i)
axiomatic, based on Wightman's axioms and Wightman's positive
definiteness condition, (ii) current algebra and (iii) path--integral,
is to use the test functions from ${\cal S}_0(\mathbb{R}^{\,2})$ [1].
In fact, in vacuum expectation values defined in current algebra and
path--integral approach the contribution of the zero--mode collective
configuration of a free massless (pseudo)scalar field can be removed
without influence on the evolution of relative motion of the
system. Therefore, from a physical point of view the definition of
Wightman's observables on the class of test functions from ${\cal
S}_0(\mathbb{R}^{\,2})$, suppressing a measurement of the collective
zero--mode, describing a shift of a free massless (pseudo)scalar
field, is well--motivated [1].

The canonical quantum field theory of the free massless (pseudo)scalar
field $\vartheta(x)$ without infrared divergences, formulated in [6],
is well defined on the class of the test functions from ${\cal
S}_0(\mathbb{R}^{\,2})$. This quantum field theory solves the problem
of infrared divergences of the Wightman functions and describes the
bosonized version of the massless Thirring model with fermion fields
quantized in the chirally broken phase. The chirally broken phase of
the massless Thirring model is characterized by a fermion condensate
[10], the non--zero value of which is caused by the non--vanishing
spontaneous magnetization, ${\cal M} = 1$, in the quantum field theory
of a free massless (pseudo)scalar field without infrared divergences
[6].

Recently [30] we have shown that the boson field representation for
the massless Thirring fermion fields, suggested by Morchio, Pierotti
and Strocchi [14,31], agrees fully with the existence of the chirally
broken phase in the massless Thirring model and the fermion
condensation. Moreover, such a representation satisfies the constant
of motion for the massless Thirring model which we have found in [10].

\section*{Acknowledgement}

\hspace{0.2in} We are grateful to our numerous referees who promoted
us to analyse Coleman's results [5,6,16,30,32].

\end{document}